%% file: main.tex
\documentclass[runningheads]{llncs}
\usepackage{graphicx}
\usepackage{csquotes}
\usepackage{multirow}
\usepackage{mathtools}
\usepackage{makecell}
\usepackage[usenames,dvipsnames,svgnames,table]{xcolor}

\usepackage{subcaption}
\usepackage{xspace}
\usepackage{amssymb, amsmath}
\makeatletter
\DeclareRobustCommand\onedot{\futurelet\@let@token\@onedot}
\def\@onedot{\ifx\@let@token.\else.\null\fi\xspace}
\def\eg{\emph{e.g}\onedot} 

\def\ie{\emph{i.e}\onedot}

\def\etal{\emph{et al}\onedot}
\makeatother
\DeclareMathOperator{\Loss}{\mathcal{L}}
\DeclareMathOperator*{\R}{\mathbb{R}}

\renewcommand{\vec}[1]{{\mathbf #1}}

\usepackage{hyperref}
\hypersetup{
    colorlinks=true,
    linkcolor=blue,
    filecolor=magenta,      
    urlcolor=cyan,
}

\begin{document}
\title{Reciprocal Adversarial Learning for Brain Tumor Segmentation: A Solution to BraTS Challenge 2021 Segmentation Task}
%
%
\author{Himashi Peiris 
\inst{1}, Zhaolin Chen 
\inst{1,2}, Gary Egan 
\inst{2}, Mehrtash Harandi 
\inst{1}}
\authorrunning{H. Peiris et al.}
%
\institute{Department of Electrical and Computer Systems Engineering, Monash University, Melbourne, Australia. \and Monash Biomedical Imaging (MBI), Monash University, Melbourne, Australia. \\
\email{\{Edirisinghe.Peiris, Zhaolin.Chen, Gary.Egan, Mehrtash.Harandi\}@monash.edu}}
\maketitle              
%

%
%

\begin{abstract}
This paper proposes an adversarial learning based training approach for brain tumor segmentation task.  In this concept,  the  3D  segmentation network learns from dual reciprocal adversarial learning approaches. To enhance the generalization across the segmentation predictions and to make the segmentation network robust, we adhere to the Virtual Adversarial Training approach by generating more adversarial examples via adding some noise on original patient data. By incorporating a critic that acts as a quantitative subjective referee, the segmentation network learns from the uncertainty information associated with segmentation results.  We trained and evaluated network architecture on the RSNA-ASNR-MICCAI BraTS 2021 dataset.  Our performance on the online validation dataset is as follows: Dice Similarity Score of 81.38\%, 90.77\% and 85.39\%; Hausdorff Distance (95\%) of 21.83 mm, 5.37 mm, 8.56 mm for the enhancing tumor, whole tumor and tumor core, respectively. Similarly, our approach achieved a Dice Similarity Score of 84.55\%, 90.46\% and 85.30\%, as well as Hausdorff Distance (95\%) of 13.48 mm, 6.32 mm and 16.98 mm on the final test dataset. Overall, our proposed approach yielded better performance in segmentation accuracy for each tumor sub-region. \href{https://github.com/himashi92/vizviva_brats_2021}{Our code implementation is publicly available}.

\end{abstract}
\input{sec_introduction}

\input{sec_related_work}
\input{sec_method}
\input{sec_experiments}

\input{sec_conclusion}

\bibliographystyle{splncs04}
\bibliography{references}

\end{document}

%% file: sec_introduction.tex
\section{Introduction}
\label{sec:introduction}
Segmentation accuracy on boundaries is essential in medical image segmentation as it is crucial for many clinical applications, such as treatment planning, disease diagnosis and image guided intervention to name a few. Tremendous progress in deep learning algorithms in dense pixel level prediction tasks has recently drawn attention on implementing automatic segmentation applications for brain tumor/giloma segmentation. Gliomas considered as the most common brain tumor variant in adults. Diagnosing High-Grade Gliomas (HGG) in early phases which are more malignant (since they usually grow fast and frequently destroy healthy brain tissue) is essential for treatment planning. On the other hand Low-Grade Gliomas (LGG) are slower growing tumors which can be cured if it is diagnosed in early phases. However, segmenting tumor sub regions from various medical images modalities (\eg, MRI and CT) is a monotonous process which is time consuming and subjective. Medical Imaging analysis is carried out by radiologists and this manual process is tedious since the volumes are hefty in size and contains heterogeneous ambiguous sub-regions (\ie. edema, active tumor structures, necrotic components, and non-enhancing gross abnormality). In particular, medical image segmentation plays a cornerstone role in computer aided diagnosis. With the recent development in computer vision algorithms in deep learning, there has been many discoveries on automatic medical image segmentation. Multi-modal brain tumor segmentation challenge (BraTS) has been one of the platforms for many discoveries for many years. During the last decade, variants of Fully convolutional networks (FCN) and Convolutional Neural Network (CNN) based architectures have shown convincing performance in previous BraTS and other segmentation challenges. Recent developments in volumetric medical image segmentation networks like 3D-Unet~\cite{cciccek20163d} and V-Net~\cite{milletari2016v} has been widely used with medical image modalities since these networks produce predictions for different planes(\ie axial (divides the body into top and bottom halves), coronal (perpendicular), and sagittal (midline of the body)). 

The main limitation of implementing and training these volumetric neural network architectures is out-of-memory (OOM) issues and extending these architectures are not feasible due to computational resource constraints. Many researchers have shown that, with a carefully crafted pre-processing, training and inference procedure, segmentation accuracy of 3D-UNet can improve further. By considering those factors like OOM issues, resource limitations, inference time, we propose an approach to tackle these challenges and 
further improve the segmentation accuracy and training process of 3D-UNet architecture~\cite{cciccek20163d}. In summary, \textbf{our major contributions} are, 

\begin{enumerate}
\item Inspired by adversarial learning techniques, we propose two way adversarial learning to segment brain tumor sub regions in multi-modal MR images.
\vspace{0.2cm}
\item We introduce a volumetric discriminator model which can explicitly show the confidence towards the current prediction to impose a higher-order consistency measure of prediction and ground truth during training.
\vspace{0.2cm}
\item We introduce Virtual Adversarial Training (VAT) during model training to enhance the model's robustness to data artefacts.
\end{enumerate}

%% file: sec_related_work.tex
\section{Related Work}
\label{sec:related_work}

\subsection{Medical Image Segmentation}
The rapid development of deep Convolutional Neural Networks and U-shaped encoder decoder architectures have shown convincing performance in medical image segmentation. The celebrated work U-Net~\cite{ronneberger2015u} has shown a novel direction to automatic medical image segmentation as it exploits both spatial and contextual information of images which greatly affect accuracy of segmentation models. Due to the simplicity and superior performance U-Net, many variants of U-shaped architectures are constantly emerging, such as Res-UNet~\cite{xiao2018weighted}, H-Dense-UNet~\cite{li2018h}, U-Net++~\cite{zhou2018unet++} and Attention-UNet~\cite{oktay2018attention}. Later, to handle volumetric medical image segmentation models are introduced into the field of 3D medical image segmentation, such as 3D-Unet~\cite{cciccek20163d} and V-Net~\cite{milletari2016v}. 

\subsection{Adversarial Learning}
Generative Adversarial Networks(GANs)~\cite{goodfellow2014generative} by Goodfellow has been a major breakthrough in the image generation task. Inspired by GAN approach, many GAN based medical imaging applications were introduced recently including in the areas of medical image segmentation~\cite{mahmood2019deep}, reconstruction~\cite{quan2017compressed} and domain adaptation~\cite{zhang2018task}. In BraTS challenge 2020, Marco \etal proposed 3D volume-to-volume Generative Adversarial Network for segmentation of brain tumours~\cite{cirillo2020vox2vox} where the discriminator is build based on PatchGAN~\cite{isola2017image} architecture style. VAT is another adversarial learning approach which has shown tremendous performance in semi-supervised learning~\cite{miyato2018virtual}. VAT is applicable to any parametric model and it directly regularizes the output distribution by its local
sensitivity of the output with respect to input~\cite{miyato2018virtual}. 

Hence, inspired by above works, we propose min-max formulation with VAT for segmenting brain tumors in multi-modal MR images. 

%% file: sec_method.tex
\section{Methodology}
\label{sec:method}
\begin{figure*}[h]
\centering
\includegraphics[width=1\linewidth]{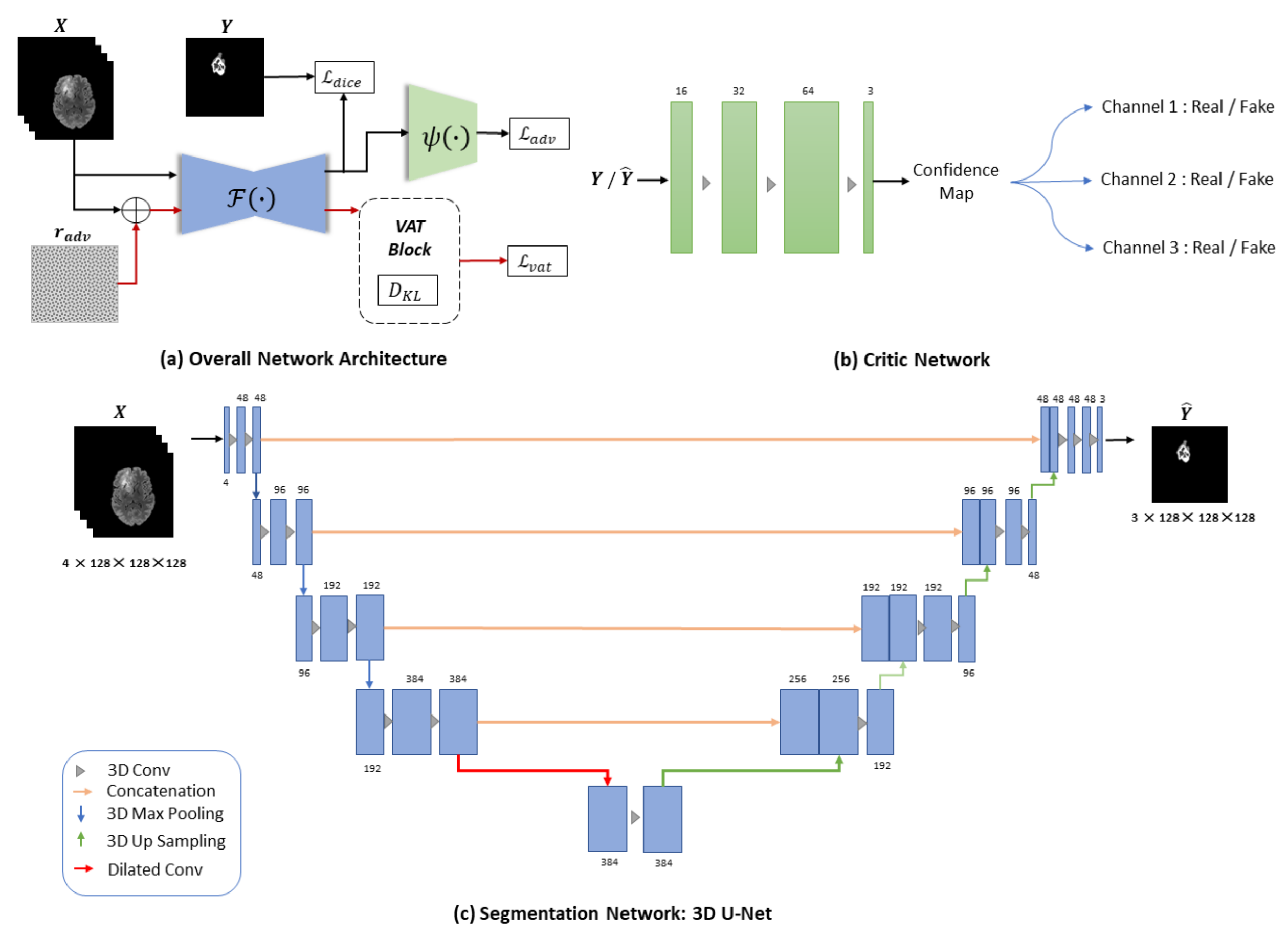} 
\caption{\textbf{Proposed Overall Network Architecture.} $\mathcal{F}(\cdot)$ and $\psi(\cdot)$ denote the Segmentation network and the Critic network. $X, Y, r_{adv}$ and $\hat{Y}$ are input data (original patient data), ground truth segmentation masks, perturbation added on input data and the prediction generated from segmentation network. Here, Critic criticizes between prediction masks and the ground truth masks to perform the min-max game by generating a pixel-wise confidence map. VAT block improves the robustness of the model against generated adversarial examples by adding perturbation that violates the virtual adversarial direction.}
\label{fig:architecture}
\end{figure*}

We start this section by providing an overview of the BraTS dataset and proposed method as shown in Fig.~\ref{fig:architecture}. Then we detail out the structure of each module and the entire training pipeline.

\subsection{Dataset}
\begin{figure*}[!htb]
\scriptsize
\tabcolsep=0.04cm
\centering
  \begin{tabular}{{c@{ } c@{ } c@{ } c@{ } c@{ }}}
    {\includegraphics[width=0.19\linewidth]{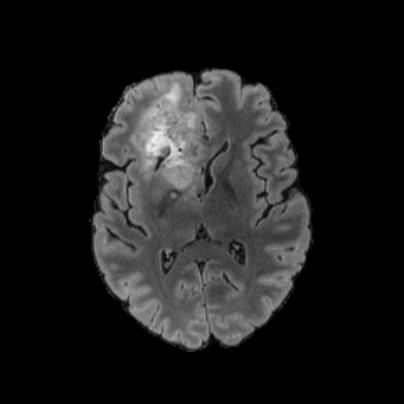}}&
    {\includegraphics[width=0.19\linewidth]{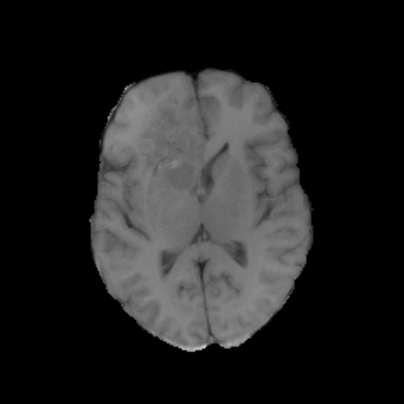}}&
    {\includegraphics[width=0.19\linewidth]{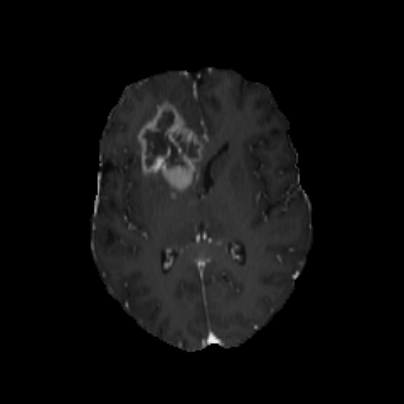}}&
    {\includegraphics[width=0.19\linewidth]{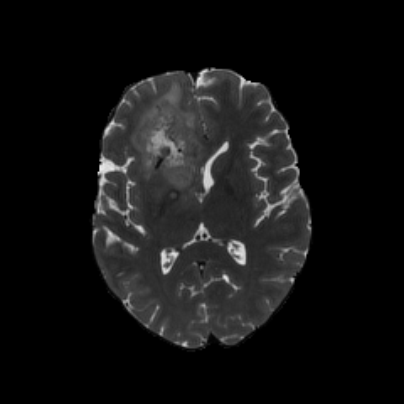}}&
    {\includegraphics[width=0.19\linewidth]{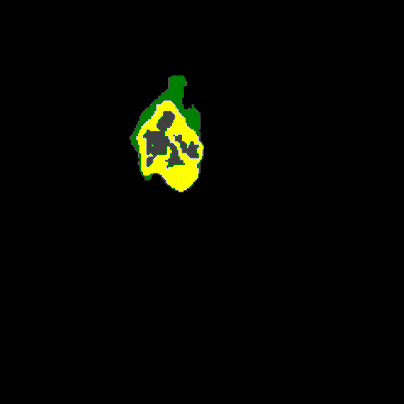}}\\
    \makecell{Flair} & \makecell{T1} &\makecell{T1CE} &\makecell{T2} &\makecell{GT} \\
  \end{tabular}
    \caption{\textbf{Visual Analysis of BraTs 2021 Training Data}. In the Ground Truth (GT) Mask, green, yellow and gray represent the peritumoral edema (ED), Enhancing Tumor (ET) and non enhancing tumor/necrotic tumor (NET/NCR), respectively. }
    \label{fig:dataset}
\end{figure*}

The Magnetic Resonance images used for the model training and evaluation are from the Multi-modal Brain tumour Segmentation Challenge (BraTS) 2021 ~\cite{baid2021rsna,menze2014multimodal,bakas2017advancing,bakas2017segmentation,bakas2017segmentation1}. The BraTS 2021 training dataset contains 1251 MR volumes of shape $240 \times 240 \times 155$. MRI is required to evaluate tumor heterogeneity. These MRI sequences are conventionally used for giloma detection: T1 weighted sequence (T1), T1-weighted contrast enhanced sequence using gadolinium contrast agents (T1Gd) (T1CE), T2 weighted sequence (T2), and Fluid attenuated inversion recovery (FLAIR) sequence. From these sequences, four distinct tumor sub-regions can be identified from MRI as: The Enhancing Tumor (ET) which corresponds to area of relative hyper-intensity in the T1CE with respect to the T1 sequence, Non Enhancing Tumor (NET), Necrotic Tumor (NCR) which are both hypo-intense in T1-Gd when compared to T1, Peritumoral Edema (ED) which is hyper-intense in FLAIR sequence. These almost homogeneous sub-regions can be clustered together to compose three semantically meaningful tumor classes as, Enhancing Tumor (ET), addition of ET, NET and NCR represents the Tumor Core (TC) region and addition of ED to TC represents the Whole Tumor (WT). MRI sequences and ground truth map with three classes are shown in Fig.~\ref{fig:dataset}.

\subsection{Problem Formulation}
Let  $\mathcal{X} = \{(\vec{X}_i,\vec{Y}_i)\}_{i=1}^m$ be a labeled set with $m$ number of samples, where each sample $(\vec{X}_i,\vec{Y}_i)$ consists of an image $\vec{X}_i \in \R^{C \times D \times H \times W}$ and its associated ground-truth segmentation mask $\vec{Y_i} \in \{0, 1, 2, 4\}^{ C \times H \times W}$. Pixels with 0,1,2 and 4 in label-map represent the background/air, Necrotic (NCR) and Non-enhancing tumor core (NET), Peritumoral Edema (ED) and Enhancing Tumor (ET). 

\subsection{Network Architecture}

The proposed network architecture consists of three modules, namely a segmentation network, a critic network and Virtual adversarial Training (VAT) block. The segmentation network (\ie, $\mathcal{F}(\cdot)$) composed of down-sampling and up-sampling layers with skip pathways, making it a U like network architecture~\cite{ronneberger2015u}. Critic is constructed as a fully convolutional adversarial network. Both networks consists 3D convolutions. The critic constructively impose the segmentation network to predict segmentation masks that are more similar to ground truth masks. The critic here, depicts Markovian PatchGAN architecture~\cite{li2016precomputed,isola2017image}. In the original work Markovian PatchGAN architecture enables producing confidence scores for prediction masks. Inspired by this, we adapt the similar approach to provide uncertainty information to the segmentation network. The VAT block generates adversarial examples, so that the segmentation network can learn to avoid making such incorrect predictions on new patient data and patient data with artefacts.

\subsection{Objective Function}
The parameters of segmentation network is defined as $\theta_G$ and the critic network is $\theta_C$. To encourage the segmentation network to yield predictions closer to the ground truth real masks by deceiving a critic network, we propose optimizing the following min-max problem:
\begin{align}
    \min_{\theta_G}\max_{\theta_C} \Loss(\theta_G; \mathcal{X})\;.
    \label{eqn:min_max_overall_loss}    
\end{align}

We propose to train the segmentation network by minimizing a the total loss function which consists of three terms:
\begin{align}
    \label{eqn:gen_loss}
    \Loss(\theta; \mathcal{X}) \coloneqq 
    \lambda_s\Loss_{dice}(\theta_G; \mathcal{X}) 
    + \lambda_v \Loss_{vat}(\theta_G; \mathcal{X}; r_{adv}) + \lambda_c\Loss_{adv}(\theta_G; \theta_C; \mathcal{X})\;, 
\end{align}
where $\Loss_{dice}$, $\Loss_{vat}$, and $\Loss_{adv}$  denote the supervised dice loss, the virtual adversarial training loss and the critic loss respectively. Furthermore, $\lambda_{\mathrm{s}}, \lambda_{\mathrm{v}}, \lambda_{\mathrm{c}} > 0$ are hyper-parameters of the algorithm, controlling the contribution of each loss term. It can be seen that the supervised dice loss and vat loss are only dependent on the segmentation networks while the critic loss is defined based on the parameters of the entire model. The segmentation network works robustly and shows generalization performance as long as these parameters are defined in a reasonable range. In our experiments we set  $\lambda_{\mathrm{s}} = 1.0$, $\lambda_{\mathrm{v}} = 0.2$ and $\lambda_{\mathrm{c}} = 0.3$.

As the main loss, we use dice loss and we calculate dice loss for each class (Multi-class loss function):
\begin{align}
    \label{eqn:dice_loss}
    \Loss_{\mathrm{dice}}(\theta_G;\mathcal{X}) \hspace{-0.1ex} &= 1 - \mathbb{E}_{(\vec{X}, \vec{Y}) \sim \mathcal{X}} \Bigg[ \frac{  
     \big \langle \vec{Y} \hspace{-0.1ex}~,~ \hspace{-0.1ex} \hat{\vec{Y}} 
     \hspace{-0.2ex} \big \rangle + \epsilon} 
     {\big\|\vec{Y}\big\|_1 + \big\|\hat{\vec{Y}} \big\|_1 + \epsilon} \Bigg],
\end{align}
where we use $\langle \vec{A},\vec{B}\rangle = \sum_{i,j,k} \vec{A}[i,j,k]\vec{B}[i,j,k]$ 
, $\| \vec{A}\|_1 = \sum_{i,j,k} |\vec{A}[i,j,k]|$ and $+ \epsilon$ is the smoothing factor (set to 1 in our experiment).

VAT is an algorithm that updates the model by the weighted sum of the gradient of the regularization term which is the second loss term of our full objective function. $\Loss_{vat}$ is a non-negative function that measures the divergence between ground truth distribution and perturbed prediction distribution. Inspired by the VAT method by Takeru \etal~\cite{miyato2018virtual}, we define the divergence based Local Distributional Smoothness (LDS) as:
\begin{align}
    \label{eqn:vat_loss}
    \Loss_{\mathrm{vat}}(\theta_G;\mathcal{X}; r_{adv}) \hspace{-0.1ex} &= \mathbb{E}_{(\vec{X}, \vec{Y}) \sim \mathcal{X}} \Bigg[ \mathcal{D}_{KL}(\vec{Y} \big\| \mathcal{F}(\theta_G, \vec{X} + r_{adv}))\Bigg].
\end{align}
Minimizing $\Loss_{vat}$ improves the generalization performance of the model and makes the model more robust against the adversarial examples that violates the virtual adversarial direction. Instead of having heavy data augmentation on the dataset with images perturbed by regular deformation we use adversarial perturbation which reduces the test error~\cite{szegedy2013intriguing}. 

We denote the functionality of the critic by $\Psi:[0,1]^{H \times W} \to [0,1]^{H \times W}$ and define the normalized loss of critic for prediction distribution as:
\begin{align}
    \Loss_{adv}(\theta_G;\theta_C; \mathcal{X}) &\coloneqq \mathbb{E}_{(\vec{X}, \vec{Y}) \sim \mathcal{X}} \Bigg[- \sum_{a \in H} \sum_{b \in W} \bigg\{(1-\eta) 
    \log\Big(\psi(\vec{Y})[a,b]\Big)
     \notag \\&+ \eta\log\Big(1 - \psi(\hat{\vec{Y}})[a,b]\Big) 
    \bigg\}\Bigg]\;,
    \label{eqn:loss_critic}
\end{align}
where $\eta = 0$ if the sample is generated by the segmentation network, and $\eta = 1$ if the sample is drawn from the ground truth labels. With this adversarial loss, segmentation network tries to deceive the critic by generating predictions that are more
similar to ground truth masks holistically.

%% file: sec_experiments.tex
\section{Experiments}
\label{sec:experiments}

\subsection{Implementation Details}
The proposed model is developed in PyTorch and trained from scratch. We use modified version of 3D UNet as the segmentation network and a 3D discriminator as the critic network. In the 3D UNet, contracting path comprises five layers including bottleneck and each consisted of two 3x3x3 convolutions together with group normalization and ReLu activation. The number of feature maps in the first encoder is predefined as 48. The down-sampling layer consists a Max pooling operation with a kernel size of 2x2x2 with stride 2. Blocks of expansive path consists performs up-sampling using the trilinear interpolation followed by 3x3x3 convolution. Final layers consists a convolutional layer of a 1x1x1 kernel with 3 output channels and a sigmoid activation. Skip connections between contracting and expansive path lead to concatenation of corresponding outputs. 3D discriminator consists 4 3x3x3 convolutions with batch normalization and leaky ReLu activation function. Discriminator here is implemented, inspired by PatchGAN~\cite{isola2017image} where cubic size is 1x1x1.

\subsubsection{Image Pre-processing}
Intensities of MRI volumes are inconsistent due to various factors such as motions of patients during the examination, different manufacturers of acquisition devices, sequences and parameters used during image acquisition. To standardize all volumes, min-max scaling was performed followed by clipping intensity values. Images were then cropped to a fixed patch size of $128 \times 128 \times 128$ by removing unnecessary background pixels. 

\subsubsection{Training}
For training of segmentation network we use Adam optimizer with the learning rate of 2e-04 and for training of critic network, we use RMSProp optimizer with the learning rate of 5e-05 as momentum based methods cause instability~\cite{arjovsky2017wasserstein}. Training was done by splitting the original training dataset into training set (80\%) and test set (20\%) for 100 epochs with batch size of 2. Therefore, 1000 MR volumes are used to train the model while 251 MR volumes were used as test set.

\subsubsection{Inference}
The BraTS 2021 validation dataset contains 219 MR volumes and synapse portal conducts the evaluation. In the inference phase, the original volume re-scaled using min-max scaling followed by clipping intensity values and cropped to $240 \times 240 \times 155$ before feeding to the saved 3D UNet model. 

\subsection{Performance Evaluation}
Segmentation accuracy of three classes (\ie, ET, TC and WT) are evaluated during training and inference. Both qualitative and quantitative analysis is performed to evaluate the model accuracy.

\begin{table*}[!htb]
\small
\centering
\caption{Validation Phase Results.}
\begin{tabular}{| l | c | c | c | c | }
\hline
Class & ~\makecell{Hausdorff \\ Distance}~ & ~Dice Score~ & ~Sensitivity~ & ~Specificity~\\
\hline
Enhanced Tumor (ET) & 21.8296 &	81.3898 & 83.3949 &	99.9695 \\
Tumor Core (TC) & 8.5632 &	85.3856 &	85.0726 &	99.9745 \\
Whole Tumor (WT) & 5.3686 &	90.7654 & 92.0858 &	99.9107 \\
\hline
\end{tabular}
\label{fig:quantitative}
\end{table*}

\begin{figure*}[!htb]
\scriptsize
\tabcolsep=0.04cm
\centering
  \begin{tabular}{{c@{ }}}
    {\includegraphics[width=0.78\linewidth]{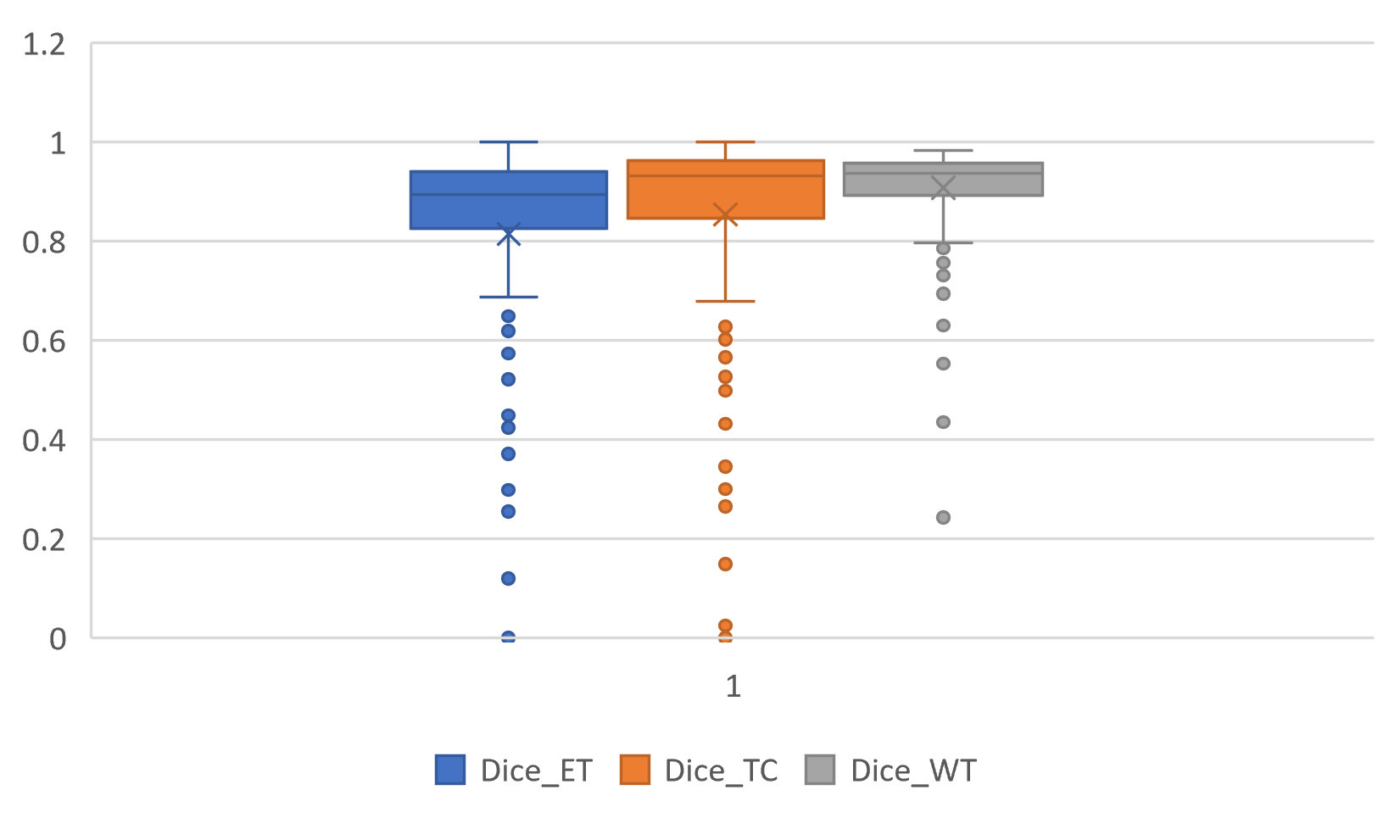}}\\
    \makecell{Dice Similarity Coefficient} \\
    {\includegraphics[width=0.78\linewidth]{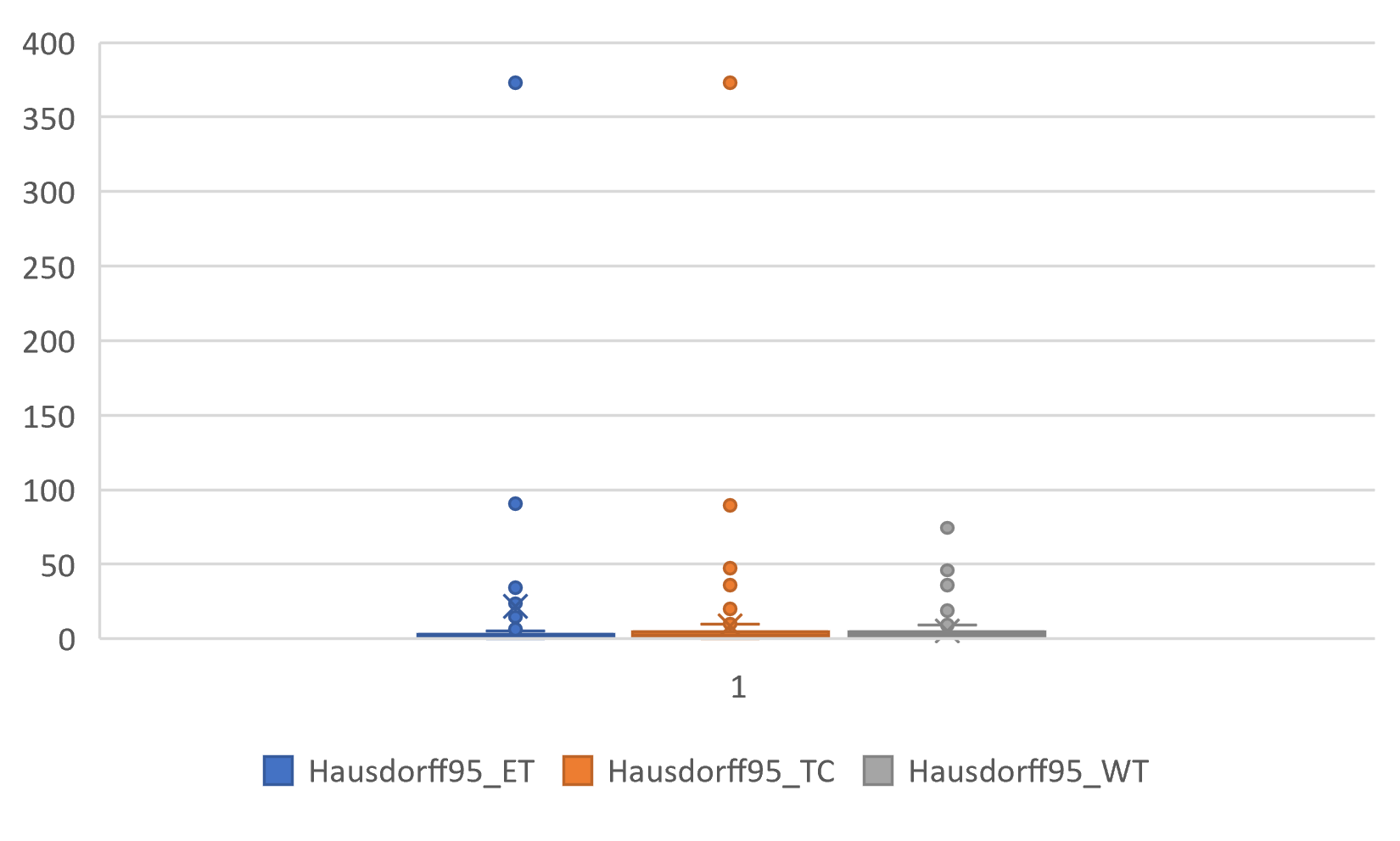}}\\
    \makecell{Hausdorff Distance} \\
  \end{tabular}
    \caption{The box and whisker plots of the distribution of the segmentation metrics for Validation Phase Results. The box-plot shows the minimum, lower quartile, median, upper quartile and maximum for each tumor class. Outliers are shown away from lower quartile.}
    \label{fig:box_plot}
\end{figure*}

\begin{figure*}[!htb]
\scriptsize
\tabcolsep=0.04cm
\centering
  \begin{tabular}{{c@{ } c@{ } c@{ }}}
    {\includegraphics[width=0.25\linewidth, trim={1cm 1cm 1cm 1cm},clip ]{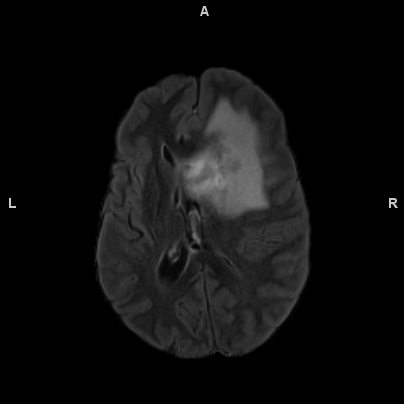}}&
    {\includegraphics[width=0.25\linewidth, trim={1cm 1cm 1cm 1cm},clip ]{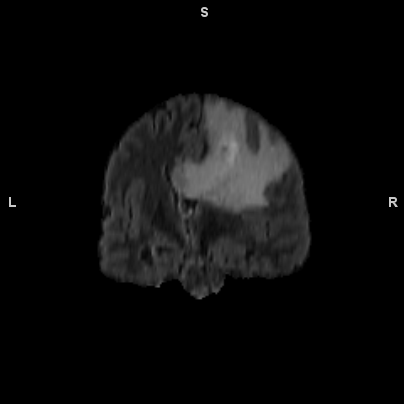}}&
    {\includegraphics[width=0.25\linewidth, trim={1cm 1cm 1cm 1cm},clip ]{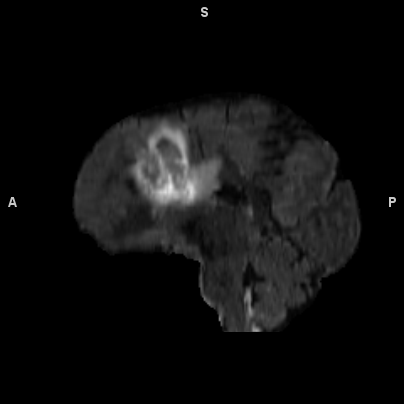}}\\
    {\includegraphics[width=0.25\linewidth, trim={1cm 1cm 1cm 1cm},clip ]{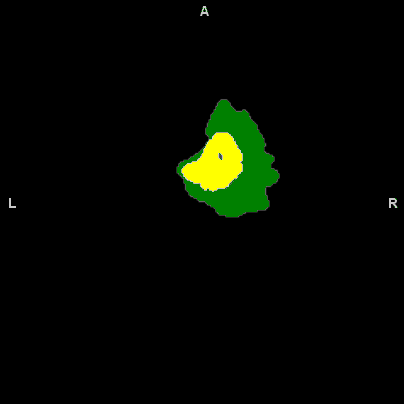}}&
    {\includegraphics[width=0.25\linewidth, trim={1cm 1cm 1cm 1cm},clip ]{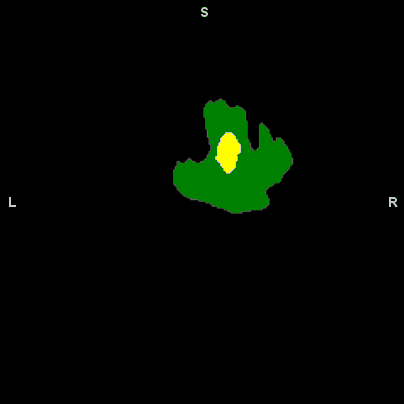}}&
    {\includegraphics[width=0.25\linewidth, trim={1cm 1cm 1cm 1cm},clip ]{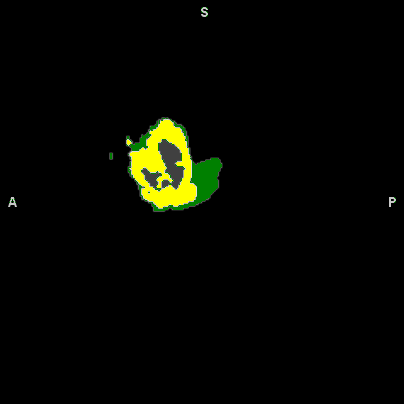}}\\
    
    \makecell{Axial View} & \makecell{Coronal View} &\makecell{Sagittal View} \\
  \end{tabular}
    \caption{Validation Phase Results for the Sample BraTS2021$\_$00190. Here, green, yellow and gray represents the Whole tumor (WT), Enhancing Tumor (ET) and Tumor Core (TC) classes respectively. (Dice (ET) = 97.2585, Dice (TC) = 99.1492, Dice (WT) = 97.5753 )}
    \label{fig:qualitative}
\end{figure*}

\subsubsection{Evaluation Matrices}
The learning model is evaluated using four matrices (1) Dice S$\o$rensen coefficient (DSC), (2) Hausdorff Distance, (3) Sensitivity and (4) Specificity. 

\subsubsection{Validation Phase Experimental Results}
The quantitative and qualitative results during validation phase for the proposed approach is shown in Table~\ref{fig:quantitative} Fig.~\ref{fig:box_plot} and Fig.~\ref{fig:qualitative}. It is noticeable that, the proposed framework helps in identifying fine predictions successfully.

\subsubsection{Testing Phase Experimental Results}
Our final evaluation results on the testing dataset are shown in Table~\ref{fig:quantitative2}. Compared to validation phase results, it can be seen that average of Dice Similarity Scores for tumor sub regions is improved during testing phase.

\begin{table*}[!htb]
\small
\centering
\caption{Testing Phase Results.}
\begin{tabular}{| l | c | c | c | c | }
\hline
Class & ~\makecell{Hausdorff \\ Distance}~ & ~Dice Score~ & ~Sensitivity~ & ~Specificity~\\
\hline
Enhanced Tumor (ET) & 13.4802 &	84.5530 & 88.0258 &	99.9680 \\
Tumor Core (TC) & 16.9814 &	85.3010 &	87.7660 &	99.9637 \\
Whole Tumor (WT) & 6.3239 &	90.4583 & 92.1467 &	99.9161 \\
\hline
\end{tabular}
\label{fig:quantitative2}
\end{table*}

%% file: sec_conclusion.tex
\section{Conclusion}
\label{sec:conclusion}
In this work, we demonstrate a simple and effective way to improve training of 3D U-Net by reciprocal adversarial learning. Our approach extends the VAT method, making the segmentation network robust to adversarial perturbations, by generating adversarial examples and adapt min-max approach adapting GAN architecture. Our experiments showed that the virtual adversarial training and uncertainty guidance help to encourage the performance of the segmentation network.

%% file: main.bbl
\begin{thebibliography}{10}
\providecommand{\url}[1]{\texttt{#1}}
\providecommand{\urlprefix}{URL }
\providecommand{\doi}[1]{https://doi.org/#1}

\bibitem{arjovsky2017wasserstein}
Arjovsky, M., Chintala, S., Bottou, L.: Wasserstein gan. arXiv preprint
  arXiv:1701.07875  (2017)

\bibitem{baid2021rsna}
Baid, U., Ghodasara, S., Bilello, M., Mohan, S., Calabrese, E., Colak, E.,
  Farahani, K., Kalpathy-Cramer, J., Kitamura, F.C., Pati, S., et~al.: The
  rsna-asnr-miccai brats 2021 benchmark on brain tumor segmentation and
  radiogenomic classification. arXiv preprint arXiv:2107.02314  (2021)

\bibitem{bakas2017segmentation}
Bakas, S., Akbari, H., Sotiras, A., Bilello, M., Rozycki, M., Kirby, J.,
  Freymann, J., Farahani, K., Davatzikos, C.: Segmentation labels and radiomic
  features for the pre-operative scans of the tcga-gbm collection. the cancer
  imaging archive. Nat Sci Data  \textbf{4},  170117 (2017)

\bibitem{bakas2017segmentation1}
Bakas, S., Akbari, H., Sotiras, A., Bilello, M., Rozycki, M., Kirby, J.,
  Freymann, J., Farahani, K., Davatzikos, C.: Segmentation labels and radiomic
  features for the pre-operative scans of the tcga-lgg collection. The cancer
  imaging archive  \textbf{286} (2017)

\bibitem{bakas2017advancing}
Bakas, S., Akbari, H., Sotiras, A., Bilello, M., Rozycki, M., Kirby, J.S.,
  Freymann, J.B., Farahani, K., Davatzikos, C.: Advancing the cancer genome
  atlas glioma mri collections with expert segmentation labels and radiomic
  features. Scientific data  \textbf{4}(1),  1--13 (2017)

\bibitem{cciccek20163d}
{\c{C}}i{\c{c}}ek, {\"O}., Abdulkadir, A., Lienkamp, S.S., Brox, T.,
  Ronneberger, O.: 3d u-net: learning dense volumetric segmentation from sparse
  annotation. In: International conference on medical image computing and
  computer-assisted intervention. pp. 424--432. Springer (2016)

\bibitem{cirillo2020vox2vox}
Cirillo, M.D., Abramian, D., Eklund, A.: Vox2vox: 3d-gan for brain tumour
  segmentation. arXiv preprint arXiv:2003.13653  (2020)

\bibitem{goodfellow2014generative}
Goodfellow, I., Pouget-Abadie, J., Mirza, M., Xu, B., Warde-Farley, D., Ozair,
  S., Courville, A., Bengio, Y.: Generative adversarial nets. In: Advances in
  neural information processing systems. pp. 2672--2680 (2014)

\bibitem{isola2017image}
Isola, P., Zhu, J.Y., Zhou, T., Efros, A.A.: Image-to-image translation with
  conditional adversarial networks. In: Proc. IEEE Conference on Computer
  Vision and Pattern Recognition (CVPR). pp. 1125--1134 (2017)

\bibitem{li2016precomputed}
Li, C., Wand, M.: Precomputed real-time texture synthesis with markovian
  generative adversarial networks. In: European conference on computer vision.
  pp. 702--716. Springer (2016)

\bibitem{li2018h}
Li, X., Chen, H., Qi, X., Dou, Q., Fu, C.W., Heng, P.A.: H-denseunet: hybrid
  densely connected unet for liver and tumor segmentation from ct volumes. IEEE
  transactions on medical imaging  \textbf{37}(12),  2663--2674 (2018)

\bibitem{mahmood2019deep}
Mahmood, F., Borders, D., Chen, R., McKay, G.N., Salimian, K.J., Baras, A.,
  Durr, N.J.: Deep adversarial training for multi-organ nuclei segmentation in
  histopathology images. IEEE Trans. on Medical Imaging  (2019)

\bibitem{menze2014multimodal}
Menze, B.H., Jakab, A., Bauer, S., Kalpathy-Cramer, J., Farahani, K., Kirby,
  J., Burren, Y., Porz, N., Slotboom, J., Wiest, R., et~al.: The multimodal
  brain tumor image segmentation benchmark (brats). IEEE transactions on
  medical imaging  \textbf{34}(10),  1993--2024 (2014)

\bibitem{milletari2016v}
Milletari, F., Navab, N., Ahmadi, S.A.: V-net: Fully convolutional neural
  networks for volumetric medical image segmentation. In: 2016 fourth
  international conference on 3D vision (3DV). pp. 565--571. IEEE (2016)

\bibitem{miyato2018virtual}
Miyato, T., Maeda, S.i., Koyama, M., Ishii, S.: Virtual adversarial training: a
  regularization method for supervised and semi-supervised learning. IEEE
  Trans. on Pattern Analysis and Machine Intelligence  \textbf{41}(8),
  1979--1993 (2018)

\bibitem{oktay2018attention}
Oktay, O., Schlemper, J., Folgoc, L.L., Lee, M., Heinrich, M., Misawa, K.,
  Mori, K., McDonagh, S., Hammerla, N.Y., Kainz, B., et~al.: Attention u-net:
  Learning where to look for the pancreas. arXiv preprint arXiv:1804.03999
  (2018)

\bibitem{quan2017compressed}
Quan, T.M., Nguyen-Duc, T., Jeong, W.K.: Compressed sensing mri reconstruction
  with cyclic loss in generative adversarial networks. arXiv preprint
  arXiv:1709.00753  (2017)

\bibitem{ronneberger2015u}
Ronneberger, O., Fischer, P., Brox, T.: U-net: Convolutional networks for
  biomedical image segmentation. In: Proc. Int, Conference on Medical Image
  Computing and Computer-Assisted Intervention (MICCAI). pp. 234--241. Springer
  (2015)

\bibitem{szegedy2013intriguing}
Szegedy, C., Zaremba, W., Sutskever, I., Bruna, J., Erhan, D., Goodfellow, I.,
  Fergus, R.: Intriguing properties of neural networks. arXiv preprint
  arXiv:1312.6199  (2013)

\bibitem{xiao2018weighted}
Xiao, X., Lian, S., Luo, Z., Li, S.: Weighted res-unet for high-quality retina
  vessel segmentation. In: 2018 9th international conference on information
  technology in medicine and education (ITME). pp. 327--331. IEEE (2018)

\bibitem{zhang2018task}
Zhang, Y., Miao, S., Mansi, T., Liao, R.: Task driven generative modeling for
  unsupervised domain adaptation: Application to x-ray image segmentation. In:
  International Conference on Medical Image Computing and Computer-Assisted
  Intervention. pp. 599--607. Springer (2018)

\bibitem{zhou2018unet++}
Zhou, Z., Siddiquee, M.M.R., Tajbakhsh, N., Liang, J.: Unet++: A nested u-net
  architecture for medical image segmentation. In: Deep Learning in Medical
  Image Analysis and Multimodal Learning for Clinical Decision Support, pp.
  3--11. Springer (2018)

\end{thebibliography}
